
\documentstyle{amsppt}
\magnification=1200
\document
\NoBlackBoxes

\define\C{\Bbb C}
\define\R{\Bbb R}
\define\Z{\Bbb Z}
\define\G{GL_n(\Bbb C)}
\define\bG{{LGL_n}(\C)}
\define\cG{\overline{ GL_n}(\C)}

\def\lG{\tilde\Cal L}
\def\d{\overline{\partial}}
\def\aG{\Cal A}
\def\lGa{\Cal L}

\def\g{\frak g}
\def\E{\Cal E}
\def\A{\Cal A}

\centerline{\bf Affine Gelfand-Dickey brackets and holomorphic
vector bundles}

\vskip .20in
\centerline{\bf Pavel I. Etingof and Boris A. Khesin}
\vskip .15in
\centerline{Yale University}
\centerline{Department of Mathematics}
\centerline{New Haven, CT 06520 USA}
\centerline{e-mail: etingof \@ math.yale.edu}
\centerline{khesin \@ math.yale.edu}
\vskip .10in
\centerline{hep-th 9312123}
\baselineskip 12pt

\vskip .1in
\centerline{\bf Abstract}
\vskip .1in

We define the (second)
Adler-Gelfand-Dickey Poisson structure on
differential operators over an elliptic curve
and classify symplectic leaves of this structure.
This problem turns out to be equivalent to
classification of coadjoint orbits for double loop algebras,
conjugacy classes in loop groups, and holomorphic vector bundles
over the elliptic curve. We show that symplectic leaves
have a finite but (unlike the traditional case of operators on the
circle) arbitrarily large codimension, and compute it
explicitly.

\centerline{\bf Introduction}
\vskip .1in

In the seventies M.Adler\cite{A},
I.M.Gelfand and L.A.Dickey \cite{GD} discovered a natural
Poisson structure on the space of $n$-th order differential operators on the
circle with highest coefficient 1 which is now called the (second)
Gelfand-Dickey bracket. This bracket arises in the theory
of nonlinear integrable equations under various names
($n$KdV-structure, classical $W_n$-algebra). B.L.Feigin proposed
to consider and study symplectic leaves for the Gelfand-Dickey
bracket -- a problem motivated by the fact that for $n=2$
these symplectic leaves are orbits of coadjoint representation
of the Virasoro algebra. A classification of
 symplectic leaves for the Gelfand-Dickey bracket
and a description of their adjacency were given in \cite{OK}.
It turned out that locally
 symplectic leaves are labeled by
one of the following:

1) conjugacy classes in
the group $GL_n$;

2) orbits of
the coadjoint representation of the affine Lie algebra
$\widehat{\frak{gl}_n}$;

3) equivalence classes of flat vector bundles on the circle of rank
$n$ (these three things are in one-to-one correspondence).

Moreover, adjacency of symplectic leaves is the same as
that for conjugacy classes, orbits and vector bundles.

Finally, the
codimension of a symplectic leaf is equal to
any of the following:

1) the dimension of the centralizer of the corresponding conjugacy class;

2) the codimension of the corresponding coadjoint orbit;

3) the dimension of the space of flat global sections of the
bundle of endomorphisms
of the corresponding flat vector bundle.

In Section 1 of
this paper we define an ``affine'' analogue of the Gelfand-Dickey
bracket. It is realized on the space of $n$-th order differential
operators on an elliptic curve which are polynomials in $\d$ with
smooth coefficients and highest coefficient 1.
The reason to consider such brackets is a search for an appropriate
two-dimensional counterpart of the theory of affine Lie algebras.
One can show that the ``affine'' analogue of the Drinfeld-Sokolov reduction
\cite{DS} sends the linear Poisson bracket on the double
loop algebra (cf.\cite {EF}) into the quadratic Gelfand-Dickey bracket
on the space of differential operators on the elliptic curve.

The main goal of the
paper is to classify and study the symplectic leaves of the affine
Gelfand-Dickey bracket.
In the case $n=2$, the problem of classification of symplectic
leaves coincides with the problem of classification of
orbits of the coadjoint representation of the complex
Virasoro algebra defined in \cite{EF} -- the Lie algebra of
pairs $(f,a)$ where $f$ is a smooth function on an elliptic curve $M$ and
$a$ is a complex number, with the commutation law
$[(f,a)(g,b)]=\bigl(f\d g-g\d f,\int_Mf\d^3g\bigr)$.

In Section 2
we show that
locally symplectic leaves of this bracket are labeled by

1) Conjugacy classes for the action of the loop group $L\G$
on the semidirect product of $\Bbb
C^*\ltimes LGL_n(\Bbb C)_0$ (where $LGL_n(\C)_0$
denotes the connected component of
the identity in the group $L\G$);

2) orbits of the coadjoint representation of the ``double''' affine
Lie algebra -- a central extension of the Lie algebra of
$\frak{gl}_n$-valued smooth functions on the elliptic curve\cite{EF};

3) equivalence classes of holomorphic vector bundles of rank $n$
and degree zero
on the elliptic curve (as before, these three things are in one-to-one
correspondence).

Since holomorphic vector bundles over an elliptic curve
are completely classified \cite{At}, this result gives
a good description of symplectic leaves.

In Section 3 we show that,
similarly to the classical case, adjacency of symplectic leaves
in the affine case
is the same as for conjugacy classes, orbits and vector bundles, and
that the codimension of a symplectic
leaf is equal to

1) the dimension of the centralizer of the corresponding conjugacy
class;

2) the codimesion of the corresponding coadjoint orbit;

3) the dimension of the space of holomorphic sections of
the bundle of endomorphisms of the corresponding holomorphic vector bundle.

In particular, this implies that in the affine case
the codimension of a symplectic leaf, though always finite,
 can be arbitrarily large, unlike the finite dimensional case, in
which it is bounded from above by $\text{dim}GL_n=n^2$.

These results constitute a two dimensional (or affine) counterpart of the
results of \cite{OK} for Gelfand-Dickey brackets.
Similarly to the non-affine case, they can be generalized to
other classical Lie groups -- $SL_n$, $Sp_{2n}$, $SO_{2n+1}$ (see \cite{OK}).

The key tool in the study of Gelfand-Dickey brackets is the
notion of monodromy of a differential operator. For the case of an
elliptic curve the monodromy is a conjugacy
class in the affine $GL_n$ (more precisely,
in the one-dimensional extension $\C^*\ltimes LGL_n(\C)_0$
of the loop group of $GL_n$). This justifies the name
``affine Gelfand-Dickey bracket''.

In Section 4 of the paper we discuss the question whether
the map assigning an equivalence class of vector bundles
to a symplectic leaf is surjective.
This question is equivalent to the question whether
any monodromy can be realized by an $n$-th order
differential operator.
For the usual Gelfand-Dickey bracket the answer to this question is
positive (it follows, for example, from the results of M.Shapiro \cite{S}).
We prove that the answer is positive in the
affine case as well, and describe an explicit realization for $n=2$ using
Atiyah's classification of vector bundles over an elliptic curve.

In Section 5 we discuss the
global structure of
the fibration of the space of differential operators by symplectic
leaves, which in the classical case is defined geometrically
by homotopy classification of quasiperiodic nonflattening curves
on a real projective space \cite{O,OK,KS}.
It turns out that the problem of counting symplectic leaves
of the affine $GL_2$-Gelfand-Dickey
bracket corresponding to the trivial rank 2 vector bundle
leads to a nice topological problem
of classification of nowhere holomorphic maps from an elliptic curve
to the complex projective line up to homotopy. In the $GL_n$ case we
encounter the problem of homotopy
classification of maps $f$ from an elliptic curve to $\Bbb CP^{n-1}$
such that the vectors $\d f,...,\d^{n-1}f$ are everywhere linearly
independent. These maps
are the affine counterparts of
nonflattening curves in $\Bbb RP^{n-1}$. At the moment a complete solution of
this problem (even in the $GL_2$-case) is unknown to the authors.

\vskip .1in
\centerline{\bf Acknowledgements}
\vskip .1in

The authors are grateful to V.Arnold, I.Frenkel, and R.Montgomery
for useful remarks.

\vskip .1in
\centerline{\bf 1. Gelfand-Dickey brackets.}
\vskip .1in

We start by recalling the definition of the Gelfand-Dickey structures
(see \cite{A,GD,DS}).

Let $M$ be a compact smooth orientable closed manifold,
$k=\Bbb R\text{ or }\Bbb C$, $C^{\infty}(M,k)$ be the algebra
of smooth $k$-valued functions on $M$, $\omega$ be a volume
form on $M$. Let $D$ be a differential operator on
$C^{\infty}(M,k)$ such that $\int_M(Df)\omega=0$ and
$D(fg)=(Df)g+f(Dg)$
for any $f,g\in C^{\infty}(M,k)$.

Define the vector space $\tilde\Cal L$
as follows:
$$
\lG=\{P=\sum_{m=0}^{n-1}u_{m+1}D^{m}|u_m\in C^{\infty}(M,k)\}.\tag 1.1
$$

To realize the dual space to $\lG$, we need to introduce
pseudodifferential symbols. They are formal expressions of the
form $\sum_{m=m_0}^{\infty}a_mD^{-m}$, $m_0\in \Bbb Z$, $a_m\in
C^{\infty}(M,k)$. It is known
that such symbols form an associative algebra: two symbols $A,B$
can be multiplied with the help of the rules $D\circ f=f\circ D +D f$,
$D^{-1}\circ f=f\circ
D^{-1}-f'\circ D^{-2}+f''\circ D^{-3}-...$,
for any $f\in C^{\infty}(M,k)$.

We realize (the regular part of) the dual space to $\lG$ as follows:
$$
\Cal A=\{A=\sum_{m=1}^{n}a_mD^{-m}|a_m\in C^{\infty}(M,k)\},\tag 1.2
$$
and the pairing $\lG\otimes \aG\to k$ is given by the formula
$$
<P,A>=\int_{M}\text{Res}(PA)\omega,\tag 1.3
$$
where $\text{Res}(X)$ is the coefficient to $D^{-1}$ in a
pseudodifferential operator $X$.
It is clear that any regular
linear functional on $\lG$ has this form.

Note that $\text{Res}(PA-AP)=Df$, where $f$ is some
function on $M$, which implies that
$\int_{M}\text{Res}(PA)\omega=\int_{M}\text{Res}(AP)\omega$.

Let $\lGa$ be the affine space of all operators of the form
$L=D^n+P$, $P\in \lG$. Clearly, the tangent space to $\lGa$ at any
point is naturally identified with $\lG$.

Following Adler, Gelfand and Dickey, let us assign a vector field
$V_A$
on $\lGa$ to every regular linear functional $A$ on $\lG$.
Its value at a point $L\in\lGa$ is:
$$
V_{A}(L)=L(AL)_+-(LA)_+L,\tag 1.4
$$
where $X_+$ denotes the differential part of $X$.

Let $\Cal C$ denote the algebra of smooth functions on $\lGa$ for $k=\Bbb
R$, and the algebra of holomorphic functions on $\lGa$ for $k=\Bbb C$.
Then assignment (1.4) allows one to define a Poisson bracket on $\Cal C$:
$$
\{ f,g\}(L)=<dg\mid_L,V_{df\mid_L}(L)>.\tag 1.5
$$
Let us call this bracket {\it the Gelfand-Dickey (GD) bracket}.
It equips $\lGa$ with a structure of a Poisson manifold.

Let us now define symplectic leaves of the GD bracket and their codimensions.

Let $L\in\lGa$. A vector $v\in T_L\lGa=\lG$ is called a Hamiltonian vector
if there exists $A\in\aG$ such that $v=V_A(L)$.

 Define the symplectic leaf $\Cal O_L$ to be the set of
all points $L'\in \lGa$ such that there exists a smooth curve
$\gamma:[0,1]\to\lGa$ such that $\gamma(0)=L$, $\gamma(1)=L'$,
and $\frac{d\gamma}{dt}$ is a Hamiltonian vector for any $t$.
It is clear that two symplectic leaves are either disjoint or
identical. Therefore, the space $\lGa$ becomes a disjoint union
of symplectic leaves.

The tangent space $T_L\Cal O_L\subset \lG$
to the symplectic leaf $\Cal O_L$ at $L$ is
obviously the space of all Hamiltonian vectors at $L$. Define the
codimension of $\Cal O_L$ to be the codimension of this tangent space
in $\lG$.
This definition makes sense because
the codimension of a symplectic leaf is the same at all its points.
\vskip .1in

We will be concerned with
the following two special cases of GD brackets.

\proclaim{Main definition}

{\bf Case 1. }
 $M=S^1$, $k=\Bbb  R$ or $\Bbb C$, $D=\frac{d}{dx}$, $\omega=dx$.
The GD bracket corresponding to this situation is called
the $GL_n(k)$-GD bracket \cite{GD}.

{\bf Case 2. }$M$ is a nondegenerate elliptic curve over $\Bbb C$:
$M=\Bbb C/\Gamma$, where $\Gamma$ is
a lattice generated by $1$ and $\tau$, where $\text{Im }\tau>0$, $k=\Bbb C$,
$D=\d=\frac{\partial}{\partial \bar z}=
\frac{1}{2}(\frac{\partial}{\partial x}+i\frac{\partial}{\partial y})$,
where $z=x+iy$ is the standard complex coordinate on $\Bbb C$,
$\omega =\frac{i}{2}dz\wedge d\bar z$.
The space $\lGa$ consists of differential operators
$\d^n+\sum_{j=0}^{n-1}u_{j+1}(z,\bar z)\d^j$, where $u_i\in C^{\infty}
(\C/\Gamma)$. We call the GD bracket corresponding to
this case the {\it affine $GL_n$-GD bracket}.
\endproclaim

Symplectic leaves of the $GL_n$-GD bracket are described in\cite{OK}.
In this paper a similar description is given for symplectic leaves
of the affine $GL_n$-GD bracket. To emphasize the parallel between the
non-affine and affine theories, we give an exposition of both of them,
marking definitions and statements from the non-affine theory by the
letter $A$ and from the affine theory by the letter $B$.

\newpage

\centerline{\bf 2. Local classification of symplectic leaves}
\vskip .1in

\proclaim{Definition 1AB} Let $\bold f=(f_1,...,f_n)$ be a smooth $k^n$-valued
function on some covering of $M$ ($k=\R$ or $\C$).
The matrix-valued function $W(\bold f)=(w_{ij})$,
$w_{ij}=D^{i-1}f_j$ is called the Wronski matrix of $\bold f$.
\endproclaim

\proclaim{Proposition 1A} Let $L$ be a differential operator
of the form
$L=\frac{d^n}{dx^n}+\sum_{j=o}^{n-1}u_{j+1}\frac{d^{j}}{dx^{j}}$,
$u_j\in C^{\infty}(S^1,k)$.
Then:

(i)  there exists a set of $n$ solutions $\bold f=(f_1,...,f_n)$
of the equation $L\phi=0$ belonging to
$C^{\infty}(\Bbb R,k)$ whose Wronski matrix is everywhere
nondegenerate (here $\Bbb R$ is regarded as a cover of $S^1$);

(ii) if $\tilde{\bold f}=(\tilde f_1,...,\tilde f_n)$ is another set of
solutions
satisfying (i) then there exists a unique matrix $R\in GL_n(k)$
such that $\tilde \bold f=\bold fR$;

(iii)  if $\bold f=
(f_1,...,f_n)$ is any set of smooth $k$-valued functions on the
real line such that its Wronski matrix is everywhere nondegenerate,
and if $\bold f(x+1)=\sum_{i=1}^n\bold f(x)R$ for some
$R\in GL_n(k)$, then there
exists a unique differential operator
$L=\frac{d^n}{dx^n}+\sum_{j=0}^{n-1}u_{j+1}\frac{d^j}{dx^j}$ with
periodic coefficients such that
$Lf_i=0$ for all $i$.
\endproclaim

\demo{Proof} This is a standard statement
from the theory of ordinary differential equations.
$\square$\enddemo

Let $\Sigma=\Bbb C/\Z$ be a cylinder. This cylinder has a natural
structure of an abelian group, is equivalent to
$\C^*$ as a complex manifold, and naturally covers
the elliptic curve $M=\C/(\Z\oplus\tau\Z)$.
  From now on we do not
make a distinction between $\Sigma$ and $\C^*$.

Before we formulate the affine analogue of Proposition 1A, we need to
define loop groups.
We will need three versions of a loop group for $\G$:

\proclaim{Notation}
$L\G$ is the group of holomorphic $\G$-valued functions on $\Sigma$.
$L\G_0$ is the connected component of identity in $L\G$.
$\cG$ is the semidirect product
$\Sigma\ltimes \bG_0$, where $\Sigma$ acts on
$\bG_0$ by $(z\circ g)(w)=g(w+z)$.
\endproclaim

The group $\cG$ should be regarded as
the group of pairs $(g(\cdot),\tau)$, $g\in L\G_0$, $\tau\in\Sigma$,
with the multiplication law $(g(z),\tau)(h(z),\theta)=(g(z)h(z+\tau),
\tau+\theta)$. It is clear that $\bG_0$ is embedded
into $\cG$ by the map $g(\cdot)\to (g(\cdot),0)$.

Consider the action of $\bG$ on $\cG$ by conjugacy. We will call
the orbits of this action {\it restricted conjugacy classes}.

\proclaim{Proposition 1B} Let $L$ be a differential operator
of the form
$L=\d^n+\sum_{j=0}^{n-1}u_{j+1}\d^{j}$,
$u_j\in C^{\infty}(M,\C)$, where $M$ is an elliptic curve.
Then:

(i)  there exists a set of $n$ solutions $\bold f=(f_1,...,f_n)$
of the equation $L\phi=0$ belonging to
$C^{\infty}(\Sigma,\C)$ whose Wronski matrix is everywhere
nondegenerate (here $\Sigma$ is regarded as a cover of $M$);

(ii) if $\tilde{\bold f}=
(\tilde f_1,...,\tilde f_n)$ is another set of solutions
satisfying (i) then there exists a unique matrix $R(z)\in L\G$
such that $\tilde \bold f=\bold fR$.

(iii)  if $\bold f=(f_1,...,f_n)$ is any set of smooth complex-valued
functions on
$\Sigma$ such that its Wronski matrix is everywhere nondegenerate,
and if $\bold f(z+\tau)=\bold f(z)R(z)$ for some $R(z)\in \bG$, then there
exists a unique differential operator
$L=\d^n+\sum_{j=0}^{n-1}u_{j+1}\d^{j}$ such that
$Lf_i=0$ for all $i$.
\endproclaim


\demo{Proof}
First of all, statements (i) and (ii) are true in a small
enough neighborhood $U_p$ of every point $p\in \Sigma$ \cite{AB}.
Let $\bold g^p=(g_1^p,....,g_n^p)$ be the corresponding sets of solutions.
By the local version of statement (ii), whenever $U_p$ and $U_q$
intersect, $g_j^p=\sum_{i=1}^ng_i^qQ_{ij}^{pq}$, where
$Q^{pq}(z)$ are holomorphic $\G$-valued functions on $U_p\cap U_q$. These
functions satisfy the conditions: $Q^{pq}Q^{qp}=1$,
$Q^{pq}Q^{qr}Q^{rp}=1$, which imply that they are
clutching transformations of some holomorphic vector bundle $E_L$ of rank
$n$ on $\Sigma$.

Since $\Sigma$ is equivalent to $\C^*$ as a complex manifold,
any holomorphic vector bundle over $\Sigma$ has to be trivial.
This, of course, applies to $E_L$, which implies that $E_L$ has
$n$ global holomorphic sections $s_1,...,s_n$ which are everywhere linearly
independent. That is to say, for every $p\in\Sigma$
there exists a holomorphic function
$S^p(z)$ on $U_p$ with values in $\G$ such that $S^p=Q^{pq}S^q$ on
$U_p\cap U_q$ for any $p,q\in\Sigma$ ($s_i$ are the columns of $S$).
Therefore, the functions $f_j^p=\sum_ig_i^pS_{ij}^p$ satisfy
the condition $f_j^p=f_j^q$ on $U_p\cap U_q$. This means, we have
a globally defined vector-function $\bold f=(f_1,...,f_n)$, such that
$f_j\mid _{U_p}=f_j^p$. Since the functions $S_{ij}^p(z)$ are
holomorphic, the functions $f_j$ satisfy the equation $Lf_j=0$.
Also, $W(\bold f)=W(\bold g^p)S^p$ in every $U_p$, which implies
$W(\bold f)$ is everywhere nondegenerate. This settles (i).

Now let $\phi$ be any smooth complex function on $\Sigma$. Consider the
column vector $\Phi=(\phi,\d\phi,...,\d^{n-1}\phi)$.
It is obvious that $\phi$ is a solution of $L\phi=0$
if and only if $\Phi$ satisfies the first order $n\times n$-matrix equation
$\d \Phi=A_L\Phi$, where $A_L$ is the Frobenius matrix corresponding
to $L$:
$$
A_L=\left(\matrix 0 & 1 & \dots & \dots & 0\\ 0 & 0 & 1 & \dots & 0
\\ \dots & \dots & \dots & \dots & \dots
\\ 0 & 0 & \dots & \dots & 1\\ -u_1 & -u_2 &\dots
& \dots & -u_n\endmatrix\right) ,\text{ i.e.  }
(A_L)_{ij}=\cases 1 & j-i=1\\ -u_{j} & i=n\\ 0 &
\text{otherwise} \endcases\tag 2.1
$$
This implies that if $\bold f=(f_1,...,f_n)$ is a set of solutions
to $L\phi=0$ then the Wronski matrix $W(\bold f)$ satisfies the equation
$$
\d W=A_LW.\tag 2.2
$$

To prove (ii), define the matrix function $R$ on $\Sigma$ by
$W(\tilde{\bold f})=W(\bold f)R$. This matrix is obviously always in
$\G$, and it is holomorphic on $\Sigma$ because both
$W(\tilde{\bold f})$ and $W(\bold f)$ satisfy the equation $\d
W=A_LW$. Thus, $R\in L\G$.

To establish (iii), for any $\bold f$ satisfying the conditions of
(iii) define the vector-function $\bold u=(u_1,...,u_n)$ on $\Sigma$
by the formula
$$
\d^n\bold f+\bold uW(\bold f)=0.\tag 2.3
$$
This vector function
exists and is unique because of the nondegeneracy of $W$.
Moreover, it is $\tau$-periodic since both $\d^n\bold f$ and $W(\bold
f)$
multiply by $R$ from the right when $z$ is replaced by $z+\tau$.
Now set $L=\d^n+\sum_{j=0}^{n-1}u_{j+1}\d^{j}$. It is obvious that (2.3)
is equivalent to the condition that $Lf_i=0$ for all $i$, which
implies (iii).
$\square$\enddemo

Propositions 1A and 1B have a simple geometric reformulation:

\proclaim{Proposition 1AB} For every
vector-function $\bold f$ with a nondegenerate Wronski matrix there
exists a unique differential operator $L_{\bold f}\in\lGa$ such that
$L_{\bold f}f_i=0$, and the assignment $\bold f\to L_{\bold f}$
is a principal fibration over $\lGa$
whose fiber is $GL_n(k)$ in Case 1 and $LGL_n(\C)$ in Case 2.
\endproclaim

\proclaim{Corollary 2AB} Let $L(t)$ be any smooth curve in $\lGa$.
Then there exists a smooth family of vector-functions
${\bold f^t}$ with a nondegenerate Wronski matrix
such that $L(t)f_i^t=0$ for all $i$ and for all $t$.
\endproclaim

\demo{Proof} This is just the statement that any path on the base of a
fiber bundle can be covered by a path on the total space.
$\square$\enddemo

Let us now define the notion of monodromy of a differential operator.

\proclaim{Definition 2A}
 Let $L$ be a differential operator
of the form
$L=\frac{d^n}{dx^n}+\sum_{j=0}^{n-1}u_{j+1}\frac{d^{j}}{dx^{j}}$,
$u_j\in C^{\infty}(\R/\Z,k)$. Let $\bold f=(f_1,...,f_n)$
be a set of solutions of the equation $L\phi=0$ belonging to
$C^{\infty}(\R,k)$ whose Wronski matrix
is everywhere nondegenerate. Let $R\in GL_n(k)$ be the matrix
such that $\bold f(x+1)=\bold f(x)R$ (it exists because of Prop. 1A (ii)).
Then the conjugacy class of $R$ in $GL_n(k)$ is called the {\it
monodromy} of $L$.
\endproclaim

Note that the matrix $R$ itself (unlike the conjugacy class of $R$,
cf. Proposition 1A (ii))
is not well defined since it relies on the choice of the set of
solutions $\bold f$.

\proclaim{Definition 2B}
 Let $L$ be a differential operator
of the form
$L=\d^n+\sum_{j=0}^{n-1}u_{j+1}\d^{j}$,
$u_j\in C^{\infty}(M,\C)$ ($M$ is an elliptic curve).
Let $\bold f=(f_1,...,f_n)$
be a set of solutions of the equation $L\phi=0$ belonging to
$C^{\infty}(\Sigma,\Bbb C)$ whose Wronski matrix
is everywhere nondegenerate. Let $R\in \bG$ be the matrix
such that $\bold f(z+\tau)=\bold f(z)R(z)$
(it exists because of Prop. 1B (ii)).
Then the restricted conjugacy class of the element $(R,\tau)$ in $\cG$ is
called the {\it
monodromy} of $L$.
\endproclaim

The reason for this definition is the following:
if $\bold g(z)=\bold f(z)Q(z)$ is another set of solutions
(i.e. $Q(z)\in LGL_n(\C)$), then $\bold g(z+\tau)=\bold g(z)\tilde
R(z)$, where $\tilde R(z)=Q^{-1}(z)R(z)Q(z+\tau)$, which corresponds
to conjugation of the element $(R,\tau)\in \cG$ by $(Q^{-1},0)$.
Since any set of solutions has the form $\bold f(z)Q(z)$, where
$Q$ is a holomorphic matrix  (Proposition 1B, part (ii)),
 monodromy is well defined, i.e.  does not depend on
the choice of $\bold f$.

Note that for differential equations on the line
there is a canonical choice of a set of solutions $\bold f$ --
the set whose Wronski matrix is the identity matrix at a fixed point
$x_0$ of the line (the fundamental system of solutions).
The notion of a fundamental system
of solutions does not have a natural analogue in two dimensions.

{\bf Remark. } Observe that in Case 2 the monodromy matrix
$R(z)$ is always in $L\G_0$. Indeed,
$\text{det}R(z)=\frac{\text{det}W(\bold f)(z+\tau)}{
\text{det}W(\bold f)(z)}$, which means that the map $z\to
\text{det}R(z)$ is homotopic to the identity:
the homotopy is $\phi_s(z)=\frac{\text{det}W(\bold f)(z+s\tau)}{
\text{det}W(\bold f)(z)}$, $s\in [0,1]$.
For a similar reason, in Case 1 if $k=\R$ then the determinant of $R$
is always positive.

\vskip .1in
Now we are ready to formulate the main theorem about the
local structure of the fibration of $\lGa$ into symplectic leaves.

\proclaim{Theorem 3AB} Let $L(t), a<t<b$ be a smooth curve in
$\lGa$. Then $L(t)$ lies inside a single symplectic leaf
if and only if the monodromy of $L(t)$ is the same for all $t$.
\endproclaim

The proof of this theorem for Case 1 was given in [OK].
Before proving Case 2, let us give a reformulation
of the isomonodromic condition in terms of vector bundles
and in terms of coadjoint orbits of double loop algebras.

Define the rank $n$ vector bundle $\E_L$ on $M$
corresponding to a differential operator $L\in\lGa$. It will be a flat
$k$-bundle in Case 1 and a holomorphic bundle in Case 2.

For every $p\in M$ let $U_p$ be the neighborhood of $p$
such that there exists a set $\bold f=(f^p_1,...,f^p_n)$ of $n$
solutions of the equation $L\phi=0$ defined in $U_p$ whose Wronski
matrix is nondegenerate in $U_p$.
Let the matrices $Q^{pq}$ (belonging to $GL_n(k)$ in Case 1 and
$LGL_n(\Bbb C)$ in Case 2) be defined by the condition $\bold
f^q=\bold f^pQ^{pq}$. Then
$Q^{pq}$ satisfy the conditions $Q^{pq}Q^{qp}=1$,
$Q^{pq}Q^{qr}Q^{rp}=1$.

\proclaim {Definition 3AB}
The vector bundle $\Cal E_L$ is the bundle on $M$ defined by the set
of transition functions $Q^{pq}$.
\endproclaim

There is another, more explicit construction of the vector bundle
$\E_L$. Let $R$ be a monodromy matrix for $L$.
Let $\hat M$ be the interval $[0,1]$ in Case 1
and the annulus $\{x+\tau y\in\Sigma|0\le y\le 1\}$ in Case 2.
Define the vector bundle $\E_L$ on $M$ as follows.
Take a trivial rank $n$ bundle over $\hat M$ and
glue the two boundaries of $\hat M$ together:
$0\sim 1$ in Case 1, $x\sim x+\tau$ in Case 2 (this will transform
$\hat M$ into $M$), identifying
the fibers over corresponding points by means of the monodromy matrix $R$.
It is easy to check that the obtained flat (holomorphic)
vector bundle over $M$ is isomorphic to $\E_L$.

Thus, global smooth sections of $\E_L$ can be realized as
quasiperiodic vector-functions on $\R$ (respectively on $\Sigma$),
i.e. as such functions $\bold f$
that $\bold f(x+1)=\bold f(x)R$ (respectively $\bold
f(z+\tau)=\bold f(z)R(z)$).

\vskip .1in
Let us now define affine and double affine Lie algebras.
 Let $\g(M)=C^{\infty}(M,{\frak
{gl}_n(k)})\oplus \Bbb C$ be
the one dimensional central extension of
$C^{\infty}(M,{\frak
{gl}_n(k)})$ by means of the cocycle
$\Omega(f,g)=\int_M\text{tr}(fDg)\omega$.
In the one-dimensional case it is the usual affine Lie algebra.
In the two-dimensional case it is the double affine algebra considered
in \cite{EF}.

It is known that the Lie algebra $\g(M)$ integrates to
a Lie group $G(M)$.
(\cite{PS} for Case 1, \cite {EF} for Case 2).
The coadjoint representation of
this group can be realized as the space of differential operators
$\lambda D+f$ ($\lambda\in k$), where $f$ is a smooth function on $M$
with values in $\frak {gl}_n(k)$, in which the action of the group
$G(M)$ reduces to the action of
$C^{\infty}(M,GL_n(k))$ by conjugation (the so called gauge
action): $g\circ(\lambda D+f)=\lambda D+Dg\cdot g^{-1}+gfg^{-1}$.
The coadjoint orbit containing the operator $\Delta=\lambda D+f$
will be denoted by $\Cal O_{\Delta}$.

The notion of monodromy for operators of the form $\lambda D+f$
($\lambda\ne 0$), where $f$ is matrix-valued,
 is analogous to that for higher order scalar operators.
For $D=d/dx$ this notion is standard; for $D=\d$, monodromy
is the restricted conjugacy class in $\cG$ of an element $(g(z),\tau)$
such that there exists a nondegenerate matrix solution
$B(z)$ of the equation $\lambda\d B+fB=0$ defined on the cylinder
$\Sigma$ and such that $B(z+\tau)=B(z)g(z)$  \cite{EF}.

Consider now the affine linear map $\Delta:\lGa\to\g(M)^*$
given by the formula $L\to D-A_L$, where $A_L$ is defined by (2.1)
(for both Case 1 and Case 2). This map takes values in the hyperplane
$\lambda=1$.

\proclaim{Proposition 4AB}
The following three conditions on two differential operators
$L_1,L_2\in\lGa$ are equivalent:

(i) $L_1$ and $L_2$ have the same monodromy;

(ii) The flat (respectively holomorphic) vector bundles
$\Cal E_{L_1}$ and $\Cal E_{L_2}$ are isomorphic.

(iii) The points $\Delta(L_1)$ and $\Delta(L_2)$ are in the same orbit
of coadjoint representation of $G(M)$.
\endproclaim

\demo{Proof} It is clear that the monodromy of the operator $L$ is the
same as the monodromy of $\Delta(L)$.

{\it Case 1. } The equivalence of (i) and (ii) is obvious;
the equivalence of (ii) and (iii) was observed in
\cite{F}, \cite{RS}, \cite{Se}.

{\it Case 2. } The equivalence of (i) and (ii) is an observation of
E.Loojienga (cf. \cite{EF}) (he observed that conjugacy classes in the extended
loop group correspond to holomorphic bundles over an elliptic curve).
The equivalence of (ii) and (iii) follows from \cite{EF}.
$\square$\enddemo

{\bf Remark. }  In Case 2 the vector bundle $\E_L$ is always of degree
zero since it is obtained from the trivial bundle on the annulus
by gluing with the help of a transition matrix $R(z)\in L\G_0$ which
is homotopic to the identity.

\demo{Proof of Theorem 3AB for Case 2} The proof
 given below follows the method of \cite{OK}.

Let $L(t)$ be a smooth curve on $\lGa$.
 Pick a smooth family of vector-functions $\bold f^t$ with a
nondegenerate Wronski matrix such that $L(t)f^t_i=0$ for all $t,i$.
This is possible because of Corollary 2AB.
Let $R^t(z)\in \bG_0$ be the monodromy matrix of this set of solutions:
it is defined by the formula $\bold f^t(z+\tau)=\bold f^t(z)R^t(z)$.

\vskip .05in
{\it If.} We must show that $L'(t)$ is a Hamiltonian vector for any
$t$.

We know that all elements $(R^t(z),\tau)$ are in the same
restricted conjugacy class
in $\cG$, i.e.  are conjugate to the same element
$(R(z),\tau)$. Therefore, $(R^t(z),\tau)$ is a smooth curve on the
restricted conjugacy class
of $(R(z),\tau)$. Since the group $\bG$ is the total space
of a principal fibration over this restricted conjugacy class
whose fiber is the centralizer of $(R(z),\tau)$ in $\bG$ (this is a
finite-dimensional complex Lie group), the curve $(R^t(z),\tau)$
can be lifted to a smooth curve $C^t(z)$ on $\bG$. In other words,
there exists a function $C^t(z)$ taking values in $L\G$
which is smooth in $t$ and satisfies the relation
$$
R^t(z)=C^t(z)R(z)(C^t)^{-1}(z+\tau).\tag 2.4
$$

Define a new vector function $\bold g^t=\bold f^tC^t$. Obviously, its
components are still solutions of $L(t)\phi=0$, and its Wronski matrix
is nondegenerate. But now we have an additional property --
the monodromy matrix of $\bold g^t$ does not depend on $t$:
$\bold g^t(z+\tau)=\bold g^t(z)R(z)$.

Let $t_0\in (a,b)$. Let $\bold g^t=\bold g+(t-t_0)\bold g'+\Cal
O((t-t_0)^2),\ t\to t_0$. Also let $L(t)=L+(t-t_0)L'+\Cal
O((t-t_0)^2),\
t\to t_0$.
Let us differentiate the relation $L(t)\bold g^t=0$ by $t$
at $t=t_0$. We get
$$
L\bold g'+L'\bold g=0.\tag 2.5
$$

 In order to show that $L'$ is a Hamiltonian
vector, we must find a pseudodifferential symbol $A$ such that
$L'=V_A(L)=L(AL)_+-(LA)_+L$. This is the same as to find an $A$ such that
$$
L\bold g'+(L(AL)_+-(LA)_+L)\bold g=0,\tag 2.6
$$
because the equation
$L\bold g'+F\bold g=0$ with respect to an $n-1$-th order differential
operator $F$ has
a unique solution: $F=\sum_{j=1}^{n}c_j\d^{j-1}$, where
$\bold c=(c_1,...,c_n)$ is equal to
 $F=-(L \bold g')W(\bold g)^{-1}$ (note that to apply a differetial operator
of order $n-1$ to a set of $n$ functions $\bold h$ is the same as to multiply
the row vector of coefficients of this operator by the Wronski matrix
 $W(\bold h)$).

Since $L\bold g=0$, equation (2.6) is equivalent to
$$
L(\bold g'+(AL)_+\bold g)=0.\tag 2.7
$$
This means that it is enough to find an $A$ such that
$$
\bold g'+(AL)_+\bold g=0.\tag 2.8
$$
That is, to find an $A$ such that
$$
(AL)_+=\sum_{j=1}^nb_{j}\d^{j-1},\tag 2.9
$$
where $\bold b=(b_1,...,b_n)$ is defined as follows:
$$
\bold b=-\bold g'W(\bold g)^{-1}.\tag 2.10
$$
Since $\bold g$ and $\bold g'$ have the same monodromy matrix,
it follows from (2.10) that $\bold b$ is doubly periodic:
$b_i\in C^{\infty}(M,\C)$.

In order to prove the existence of $A$ satisfying (2.9), it suffices
to show that the linear map $\chi:\A\to \lG$ given by $\chi(A)=(AL)_+$
is an isomorphism.
But this is obvious: the coefficients of the operator
 $(AL)_+$, have the triangular form $a_i+P_i$, where $P_i$ is a differential
polynomial in $a_1,...,a_{i-1}$, and hence the coefficients $a_i$ of
the solution of the equation
$(AL)_+=\Lambda$, $\Lambda\in\lG$, can be uniquely
determined recursively starting from $a_1$.

\vskip .05in
{\it Only if. } Differentiating the equation $L(t)\bold
f^t=0$, we get
$$
L\bold f'+L'\bold f=0.\tag 2.11
$$
(we use the shortened notation $\bold f$ for $\bold f^t$).
 We know that
$L'=V_A(L)$ for some $A$. This implies:
$$
L(\bold f'+(AL)_+\bold f)=0.\tag 2.12
$$
This means that
$$
\bold f'+(AL)_+\bold f=\bold h,\tag 2.13
$$
where $\bold h$ satisfies the equation $L\bold h=0$.

Let us show that we could have chosen $\bold f^t$ in such a way that
$\bold h=0$. Indeed, let $\bold g^t$ be another set of solutions of
$L\phi=0$ given by
$$
\bold g^t=\bold f^t(C^t)^{-1},\tag 2.14
$$
where $C^t\in L\G$. Substituting (2.14) in (2.13), we get
$$
\bold g'C+\bold gC'+(AL)_+\bold gC=\bold h\tag 2.15
$$
(here we used the shortened notation $\bold g$ for $\bold g^t$,
and $C$ for $C^t$). We want to have the relation $\bold g'+(AL)_+\bold
g=0$. This is equivalent to the relation $\bold gC'=\bold h$, or, in terms
of $\bold f$, $\bold fC^{-1}C'=\bold h$. This happens
if and only if $C^{-1}C'=W(\bold h)W(\bold f)^{-1}$, or
$C'=W(\bold h)W(\bold f)^{-1}C$. This is a first order differential
equation on $L\G$ (since $W(\bold h)W(\bold f)^{-1}$ is a holomorphic
matrix-valued function), and it has a unique solution with the initial
condition $C(t_0)=\text{Id}$.

Therefore, we may assume that $\bold h$ in (2.13) is equal to $0$.

We have
$$
\bold f'(z)=-(AL)_+\bold f(z).\tag 2.16
$$
 Changing $z$ to $z+\tau$ and using the
monodromy relation $\bold f(z+\tau)=\bold f(z)R(z)$ ($R=R^t$),
we get
$$
\bold f'(z)R(z)+\bold f(z)\frac{\partial R}{\partial t}(z)=-(AL)_+\bold
f(z)R(z),\tag 2.17
$$
which, together with (2.16), implies $\bold f(z)\frac{\partial R}
{\partial t}(z)=0$.
Therefore, $W(\bold f)\frac{\partial R}{\partial t}=0$,
which means $\frac{\partial R}{\partial t}=0$, or $R^t(z)$ is
independent of $t$. Thus, the monodromy of $L(t)$ is independent of
$t$ Q.E.D.
$\square$\enddemo

\vskip .1in
\centerline{\bf 3. Codimension and adjacency of symplectic leaves.}
\vskip .1in

\proclaim{Theorem 5AB} Let $L\in\lGa$ be a differential operator. Then
the following four numbers coincide:

(i) the codimension of the symplectic leaf $\Cal O_L$;

(ii) the dimension of the centralizer of the monodromy matrix of $L$;

(iii) the codimension of the orbit $\Cal O_{\Delta(L)}$
in the hyperplane $\lambda=1$ in the coadjoint representation
of the group $G(M)$ (see Section 2);

(iv) the dimension of the space of global sections of
the vector bundle $\text{End}(E_L)=\Cal E_L\otimes\Cal E_L^*$
(flat sections for Case 1, holomorphic sections for Case 2).
\endproclaim

{\bf Remarks.} 1. By the codimension of an orbit of the coadjoint
representation we mean the codimension (in the hyperplane $\lambda=1$)
of the tangent space to the
orbit at any point.

2. Sometimes we will call the dimension of the centralizer of
a (restricted) conjugacy class the {\it codimension} of this conjugacy
class.

3. For Case 1, it is easy to show that the number
(i)-(iv) is finite. In Case 2, it follows from algebraic geometry that (iv)
is finite, and Theorem 5AB
implies that so are (i),(ii),(iii).

{\bf Observation.} We know that symplectic leaves of the classical
(respectively, affine) GD bracket are labeled by conjugacy
classes in $GL_n(k)$ (respectively, $\cG$).
It turns out, however, that in the affine case conjugacy classes
close enough to the ``identity'' $(\text{Id},\tau)$ in
$\cG$ can be labeled by congugacy classes of the finite-dimensional
group $GL_n(\Bbb C)$. Indeed, near the ``identity'' the group $\cG$ is
identified with a region in its Lie algebra by the exponential map.
The Lie algebra of $\cG$ can be thought of as the coadjoint
representation of the affine Lie algebra $\widehat{\frak{gl}_n}$
(i.e. the space of differential operators $\lambda
\frac{d}{dz}-A(z)$). Therefore, the conjugacy classes become
coadjoint orbits for the affine Lie algebra
$\widehat{\frak{gl}_n}$, and those are enumerated by $\lambda$ and
the monodromy of the corresponding operators $\lambda
\frac{d}{dz}-A(z)$ (see \cite{F},\cite{RS}).

\demo{Proof of Theorem 5AB}

(i)=(ii).
Let $L\in \lGa$.

Let $\bold f$ be a set of solutions of $L\phi=0$ with a nondegenerate
Wronski matrix. Let $R$ be the monodromy matrix of $\bold f$:
$\bold f(x+1)=\bold f(x)R$, $R\in GL_n(k)$ (Case 1),
$\bold f(z+\tau)=\bold f(z)R(z)$, $R\in LGL_n(\C)_0$ (Case 2).

We will describe the tangent space
$T_L\Cal O_L$ as the image of a
certain operator.

Consider
the linear operator $\hat L(\bold g)=(L\bold g)W(\bold f)^{-1}$
sending the space of vector-functions $\bold g=(g_1,...,g_n)$
such that
$$
\bold g(z+\tau)=\bold g(z)R(z).\tag 3.1
$$
to the space of doubly periodic vector-functions.

{\bf Lemma. } The tangent space $T_L\Cal O_L$ is the set of all
differential operators of the form $\sum_{i=0}^{n-1}p_{i+1}D^i$, such
that the vector $\bold p=(p_1,...,p_n)$ belongs to the image of $\hat L$.

{\it Proof of the Lemma}

Applying equation (1.4) to $\bold f$, we get
$$
V_A(L)\bold f=L(AL)_+\bold f.\tag 3.2
$$
Let $V_A(L)=\sum_{i=0}^{n-1}p_{i+1}D^{i}$, and let $\bold p=(p_1,...,p_n)$.
Then (3.2) can be rewritten in the form
$$
\bold pW(\bold f)=L(AL)_+\bold f.\tag 3.3
$$
We know that $(AL)_+$ can be any differential operator
of the form $\sum_{i=0}^{n-1}b_{i+1}D^{i}$, $b_i\in C^{\infty}(M,k)$.
Therefore, the set of possible values of the expression
$(AL)_+\bold f$ is the set of all vector-functions $\bold g$ on
the cylinder satisfying (3.1).
 Indeed, (3.1) clearly must be satisfied, and whenever $\bold g$
does satisfy (3.1), one can set $\bold b=\bold gW(\bold f)^{-1}$
and get a doubly periodic vector-function.

This consideration implies that the set of possible values
of $\bold p$ is the image of the operator $\hat L$, Q.E.D.
$\square$

The Lemma shows that the set of possible values of $\bold pW(\bold f)$
is the image of the operator $L$ regarded
as an operator on the space of vector-functions $\bold g$
satisfying (3.1), i.e. on the space of smooth sections of the vector
bundle $\E_L$. The codimension of $T_L\Cal O_L$ is therefore equal
to the codimension of this image, since $W(\bold f)$ is just an
automorphism of $\lG$.

The operator $L:\Gamma(\E_L)\to \Gamma(\E_L)$ is an elliptic operator
on the circle (torus), so its index is equal to zero. Therefore,
the dimension of its kernel is equal to the codimension of its image.
Thus, it remains to compute the dimension of the kernel of $L$.

An element that undoubtedly belongs to $\text{Ker}L$ is $\bold f$.
Furthermore, any other element $\bold g$ of this kernel, according to
Proposition 1AB, can be represented in the form $\bold g=\bold fC$,
where $C$ is an $n\times n$-matrix in Case 1 and
a holomorphic $n\times n$-matrix valued function
on $\Sigma$ in Case 2.
The matrix $C$
has to satisfy the relation
$$
\gather
C=R^{-1}CR\ \text{(Case 1)}\\
C(z+\tau)=R^{-1}(z)C(z)R(z)\ \text{(Case 2)},\tag 3.4
\endgather
$$
which is equivalent to $C$ being in the Lie algebra
of the centralizer of the monodromy of $L$.
This shows that $\text{Ker}L$ is isomorphic to the Lie algebra
of the centralizer, i.e.  their dimensions are the same.

(ii)=(iv) The solutions of (3.4) are exactly the
flat (respectively holomorphic) sections of the vector
bundle $\text{End}(\E_L)=\E_L\otimes \E_L^*$, and vice versa.

(iii)=(iv) Let $\Delta=D-A\in\g(M)^*$. Then the tangent space
to the coadjoint orbit at $\Delta$, $T_{\Delta}\Cal O_{\Delta}$,
consists of vectors of the form $DX-[A,X]$, where $X$ is an arbitrary
matrix-valued function on $M$. Therefore, the codimension of the orbit
is equal to the codimension of the image of the operator
$D-\text{ad}A$ in $C^{\infty}(M,\frak{gl}_n(k))$. Since this operator
is elliptic, its index is zero, so the codimension of its image equals
the dimension of its kernel. But the kernel of this operator
consists of flat (respectively holomorphic) sections
of the bundle $\E_L\otimes \E_L^*$ and only of them.
Therefore, the dimensions of the kernel and the space of sections
coincide.
$\square$\enddemo

\proclaim{Proposition 6AB}
The codimension of every symplectic leaf (coadjoint orbit, conjugacy
class) is congruent to $n$ modulo $2$.
\endproclaim

\demo{Heuristic proof} Thanks to Theorem 5AB, it is enough to give a
proof for coadjoint orbits. Coadjoint orbits have
a natural symplectic (or holomorphic symplectic) structure -- the
Kirillov-Kostant structure. Therefore, they must all be
``even-dimensional'', i.e. their codimensions must have the same
parity. Also, the orbit corresponding to $\Delta=\d$ has codimension
$n^2$, which is congruent to $n$ modulo $2$.
Therefore, all codimensions must be congruent to $n$ modulo $2$,
Q.E.D.
$\square$\enddemo

This proof is instructive but, unfortunately, not satisfactory from
the point of view of infinite-dimensional analysis, so we
give a rigorous algebraic proof.

\demo{Rigorous proof}
{\it Case 1. } Because of Theorem 5AB, it is enough to
show that codimensions of all conjugacy classes in $GL_n(k)$ have
the same parity. This follows from the fact that all conjugacy
classes in $GL_n(k)$ are even-dimensional -- a standard fact from
linear algebra.

{\it Case 2. } Because of Theorem 5AB, Proposition 6AB is equivalent to the
assertion that for any rank $n$ holomorphic vector bundle $E$ of degree zero
over an elliptic curve $M$ the dimension of the space $H^0(M,E\otimes E^*)$
of global holomorphic sections of the bundle $E\otimes E^*$ is
congruent to $n$ modulo $2$. This assertion is a corollary of the
following Lemma.

\vskip .08in
{\bf Lemma. } Let $E$ be a holomorphic vector bundle over
an elliptic curve $M$ of rank $r$ and degree $d$.
Then $\text{dim}H^0(M,E\otimes E^*)\equiv rd+r+d\text{ mod }2$.
\vskip .08in

{\it Proof of the Lemma.} Let $V$ be a holomorphic vector bundle over
$M$ of degree $d$. Then by the Riemann-Roch theorem
$\text{dim}H^0(M,V)-\text{dim}H^1(M,V)=d$. Also, Serre's
duality tells us that $H^0(M,V^*)=H^1(M,V)^*$. Combining these two
facts, we get:
$$
\text{dim}H^0(M,V\oplus V^*)\equiv d\text{ mod }2.\tag 3.5
$$

The proof of the Lemma is by induction. For line bundles the statement
is obvious. We assume that we know the Lemma is true for bundles of
rank $l<m$. Let $E$ be a bundle of rank $m$. We consider two possibilities.

1) $E$ is indecomposable. Then a theorem of Atiyah's \cite{At} tells us
that $\text{dim}H^0(M,E\otimes E^*)$ equals the greatest common
divisor $(r,d)$ of the rank $r$ and the degree $d$ of $E$.
But $(r,d)\equiv rd+r+d\text { mod }2$, Q.E.D.

2) $E=E_1\oplus E_2$. Then
$$
H^0(M,E\otimes E^*)=H^0(M,E_1\otimes E_1^*)\oplus H^0(M,E_2\otimes E_2^*)
\oplus H^0(M,E_1\otimes E_2^*\oplus E_2\otimes E_1^*).\tag 3.6
$$
Using the assumption of induction, congruence (3.5), and the facts that
$(E_1\otimes E_2^*)^*=E_2\otimes E_1^*$ and
$\text{deg}(E_1\otimes E_2^*)=r_1d_2+r_2d_1$, we get the
congruence
$$
\text{dim}H^0(M,E\otimes E^*)\equiv (r_1d_1+r_1+d_1)+(r_2d_2+r_2+d_2)+
(r_1d_2+r_2d_1)\text{ mod }2,\tag 3.7
$$
where $r_i$ are the ranks and $d_i$ are the degrees of $E_i$.
But the right hand side of (3.7) equals to
$(r_1+r_2)(d_1+d_2)+(r_1+r_2)+(d_1+d_2)=rd+r+d$, Q.E.D.
$\square$\enddemo

Let us now discuss adjacency of symplectic leaves.

\proclaim{Definition 4AB}

(i) A symplectic leaf (coadjoint orbit) $\Cal O_1$ is called adjacent to
a symplectic leaf (coadjoint orbit) $\Cal O_2$ if there exists
a smooth curve $\gamma(t)$ in the space of differential operators
such that $\gamma(0)$ belongs to $\Cal O_1$ and $\gamma(t)$ belongs to
$\Cal O_2$ for $t\ne 0$.

(ii) A conjugacy class $C_1$ is called adjacent to a conjugacy class
$C_2$ if there exists a smooth curve $\gamma(t)$ on the group
such that $\gamma(0)\in C_1$, and $\gamma(t)\in C_2$, $t\ne 0$.

(iii) A flat (holomorphic) vector bundle $E_1$ is called adjacent
to a bundle $E_2$ if there exists an open cover $\{U_i\}$ of $M$ and
a set of transition functions $R_{ij}^t$ smooth in $t$
such that they define a bundle  isomorphic to $E_1$ at $t=0$ and to
$E_2$ at $t\ne 0$.
\endproclaim

{\bf Remarks. } 1. A symplectic leaf (coadjoint orbit) $\Cal O_1$
is adjacent to $\Cal O_2$
if and only if the closure of $\Cal O_2$ in
$C^{\infty}$-topology contains $\Cal O_1$.

2. A symplectic leaf (coadjoint orbit) $\Cal O_1$
is contained in the closure of $\Cal O_2$ if and only if
at least one point of $\Cal O_1$ is contained in this closure.

\proclaim{Proposition 7AB}
(i) Adjacency of two symplectic leaves is equivalent
to adjacency of the corresponding coadjoint orbits, conjugacy
classes, and vector bundles.

(ii) The codimension of a symplectic leaf (coadjoint orbit, conjugacy class)
 is less than the codimension of all symplectic leaves
(coadjoint orbits, conjugacy classes) adjacent to it.

(iii) Adjacency is a partial order on the set of symplectic leaves,
coadjoint orbits, conjugacy classes, and vector bunldes.
\endproclaim

\demo{Proof}
Statement (i) for Case 1 is proved in \cite{OK}
(the proof is straighforward), for Case 2 it is analogous.
Statement (ii) follows from Theorem 5AB and
the fact that in a smooth family of vector
bundles the dimension of the space of sections is
lower semicontinuous, i.e.  $\lim _{t\to t_0}\text{dim}(t)\le
\text{dim}(t_0)$. Statement (iii) follows from the fact that the
relation of adjacency introduces a (non-Hausdorff) topology
on the set of symplectic leaves (orbits etc.) in which the closure of a leaf
is the union of this leaf and all the leaves adjacent to it.
$\square$\enddemo

{\bf Remark. } More generally, one can define versal deformations of symplectic
leaves following \cite{LP},\cite{OK}. They are equivalent to
the deformations of the corresponding monodromies.

\proclaim{Definition 5AB} A symplectic leaf (coadjoint orbit,
conjugacy class) is called closed if no other symplectic
leaf (coadjoint orbit, conjugacy class) is adjacent to it.
\endproclaim

{\bf Remark. } A symplectic leaf (coadjoint orbit,
conjugacy class) is closed according to Definition 5AB
if and only if it is closed in the usual sense, i.e.  in
$C^{\infty}$-topology.

\proclaim{Corollary 8AB}
(i) In Case 1, a symplectic leaf (coadjoint orbit, conjugacy class)
is closed
 if and only if the corresponding flat vector bundle
is semisimple, i.e. a direct sum of flat line bundles.

(ii) In Case 2, a symplectic leaf (coadjoint orbit, conjugacy class)
is closed
if and only if the corresponding holomorphic vector bundle
is semisimple, i.e. a direct sum of holomorphic line bundles of degree zero.
\endproclaim

\demo{Proof}
Statement (i) follows from \cite{OK}. Statement (ii)
is proved analogously. The proof is based on the following
property: no vector bundle is adjacent to a vector bundle $E$
over an elliptic curve if and only if this bundle is a direct sum of
line bundles of degree zero.
This property follows from Atiyah's classification
of holomorphic vector bundles on an elliptic curve \cite{At}.
$\square$\enddemo

\vskip .1in
\centerline{\bf 4. Existence of differential operators with a prescribed
monodromy}
\vskip .1in

A natural question in the theory of differential equations
is: given a conjugacy class in $GL_n(k)$, does there exist
a differential operator $L\in \lGa$ whose monodromy
is this conjugacy class? In other words, is the map
assigning conjugacy classes to symplectic leaves of the
$GL_n$-Gelfand-Dickey
bracket
surjective?
The answer to this question is positive:

\proclaim{Proposition 9A} Any matrix in $GL_n(k)$ (with positive
determinant if $k=\R$) is a monodromy
matrix of an $n$-th order differential operator on the circle
with the highest coefficient 1.
\endproclaim

\demo{Proof} For $k=\R$, this proposition
is proved in \cite{S}. For $k=\C$, the proposition is obvious. Indeed,
take any matrix $R\in GL_n(\C)$, construct any vector-function
$\bold f:\R\to \C^n$ with the property $\bold f(x+1)=\bold f(x)R$.
Compute the Wronski determinant $W(x)$ of $\bold f$. This is a curve
in the complex plane. We can assume that
this curve does not go through the origin: if it does,
we can correct it by a small perturbation of $\bold f$.
Now, by virtue of Proposition 1A (iii)
there exists an $n$-th order differential equation
$L\phi=0$ with highest coefficient 1
for which $\bold f$ is a set of linearly independent solutions.
$\square$\enddemo

One may now ask if Proposition 9A can be generalized to Case 2, i.e
whether the map assigning restricted conjugacy classes in $\cG$ to symplectic
leaves of the affine Gelfand-Dickey bracket is surjective.

\proclaim{Theorem 9B} Every holomorphic vector bundle over an
elliptic curve $M$ arises as monodromy of an $n$-th order
operator $L=\d^n+\sum_{j=0}^{n-1}u_{j+1}\d^j$, $u_j\in C^{\infty}(M,\C)$.
\endproclaim

\demo{Proof} Thanks to Proposition 1B,
it suffices to prove the following statement: for any monodromy matrix
$R(z)\in LGL_n(\C)_0$ there exists a smooth vector function
$\bold f: \Sigma\to\C^n$, $\bold f=(f_1,...,f_n)$, such that
$\bold f(z+\tau)=\bold f(z)R(z)$, and the Wronslian of $\bold f$ does
not vanish on $\Sigma$.

First of all, the vector bundle on $M$ prescribed by the gluing
function $R(z)$ is topologically trivial since $R(z)$ is homotopic to
the identity. Therefore, it admits a smooth trivialization --
a smooth function $X:\Sigma\to GL_n(\C)$ such that
$X(z+\tau)=X(z)R(z)$. Let us look for the vector function
$\bold f$ in the form $\bold f=\bold gX$, $g=(g_1,...,g_n)$.
Then the monodromy condition
on $\bold f$ is equivalent to the condition that $\bold g$ is
$\tau$-periodic, i.e. that $\bold g\in C^{\infty}(M,\C^n)$.

Let $Y(z)=\d X(z)\cdot X(z)^{-1}$. This is a smooth
matrix-valued function periodic with periods $1$ and $\tau$, i.e. a
function on $M$. Consider the operator $D$ on
$C^{\infty}(M,\C^n)$ defined by $D\bold g=\d \bold g +\bold g Y$.

It is easy to check that the Wronski matrix of $\bold f$
can be written in the form $W(\bold f)=W_{D}(\bold g)X$,
where $W_{D}(\bold g)_{ij}=(D^{i-1}\bold g)_{j}$ (i.e. the lines of
$W_{D}(\bold g)$ are $\bold g$, $D\bold g$, $D^2\bold g$,...).
Therefore, our problem reduces to finding $\bold g$
such that $W_{D}(\bold g)$ is everywhere nondegenerate.
This can be done as follows.

Let $z=x+\tau y$, $x,y\in\R$. Set $g_m(z)=e^{2\pi imkx}$, $1\le m\le
n$, where $k$ is an integer. If we regard $k$ as an independent
variable, then the expression $W_{D}(\bold g)$ is a polynomial in $k$
and $e^{2\pi i kx}$ (with coefficients dependent of $z$).
The highest term in $k$ is the usual Wronskian $W(\bold g)$,
which equals $(\pi i k)^{n(n-1)/2}V_ne^{\pi ikn(n+1)x}$,
where $V_n$ is the Vandermonde determinant of $1,2,...,n$.
The absolute value of this term equals $|V_n| (\pi k)^{n(n-1)/2}$,
which grows as $k^{n(n-1)/2}$ as $k\to\infty$.
The rate of growth of the terms with lower degrees of $k$ is lower,
so for $k$ big enough (uniformly in $x,y$) the highest term will dominate.
Therefore, $W_{D}(\bold g)$ does not vanish if $k$ is big enough,
Q.E.D. $\square$

Let us describe an explicit realization of vector bundles by
differential operators for $n=2$.
 Before we do so, let us formulate Atiyah's
classification theorem for vector bundles of rank 2.

\proclaim{Atiyah's theorem}
(for rank 2 bundles)\cite{At}
Any rank 2 holomorphic vector bundle of degree zero over an elliptic curve
$M=\C/(\Z\oplus \tau\Z)$, $\tau\in C^+$,
is isomorphic to one of the following:

1) $E(a,b,m)$ ($a,b\in \C^*$, $m\in \Z$, $m\ge 0$) --
the vector bundle corresponding to the conjugacy class
of the element
$$
\left(\biggl[\matrix ae^{2\pi imz}&0\\ 0& be^{-2\pi im z}\endmatrix
\biggr],\tau\right)\tag 4.1
$$
of $\cG$.
The bundles $E(a_1,b_1,m_1)$ and $E(a_2,b_2,m_2)$
are isomorphic iff $m_1=m_2$, $a_1/a_2=q^{k_a}$,
$b_1/b_2=q^{k_b}$,
where $k_a,k_b\in \Z$, and $q=e^{2\pi i\tau}$.

2) $F(a)$, $a\in\C^*$ -- the vector
bundle corresponding to the conjugacy class
of the element
$$
\left(\biggl[\matrix a&1\\ 0& a\endmatrix
\biggr],\tau\right)\tag 4.2
$$
of $\cG$;
the bundles $F(a)$ and $F(b)$ are isomorphic iff $a/b=q^k$, $k\in\Z$.

A bundle $F(a)$ is never isomorphic to $E(a,b,m)$.
\endproclaim

Let us now realize each bundle from classes 1) and 2) by
a differential operator $L=\d^2+u_1\d+u_2$.

Observe that if a bundle $E$ is realizable by a differential
operator then so is $X\otimes E$, where $X$ is an arbitrary degree
zero line
bundle. Indeed, let $X$ correspond to the conjugacy class
of the element $(a,\tau)\in \overline{GL_1(\C)}$, $a\in \C^*$.
Let $E$ be realized by a differential operator $L$.
Then it is easy to see that $X\otimes E$ is realized
by the differential operator
$\tilde L=e^{\alpha(z-\bar
z)}\circ L\circ e^{-\alpha(z-\bar z)}$, where
$$
\alpha=\frac{\log
a}{\tau-\bar\tau}\tag 4.3
$$
 (any branch of $\log$ can be taken).

This observation implies that it is enough for us
to realize the bundles $E(a,a^{-1},m)$ and $F(1)$ by differential
operators, since all the other bundles can be obtained by tensoring them with
line bundles.

It is easy to see that the bundle $F(1)$ is realized by
the operator $L=\d^2$; the corresponding vector $\bold f$ of
solutions is $(1,y)$, where $z=x+\tau y$. The bundle
$E(a,a^{-1},0)$ is realized by the operator
$L=\d^2-\alpha^2$, where $\alpha$ is defined by (4.3)
(any nonzero value of $\log$ can be taken);
the corresponding vector $\bold f$ of
solutions is $(e^{\alpha(z-\bar z)}, e^{-\alpha(z-\bar z)})$.

It remains to realize the bundles $E(a,a^{-1},m)$ for $m>0$.

Let $z_1,...,z_m\in M$ be pairwise distinct points,
and let $\psi:M\to\C$ be a smooth function on the elliptic curve which
has the following properties:

(i) $\psi$ vanishes at $z_1,...,z_m$ and nowhere else;

(ii) in the neighborhood of $z_i$ the function $\psi$ has the form
$$
\psi(z)=|z-z_i|^2.\tag 4.4
$$

Such a function is very easy to construct: set
$$
\psi(z)=\psi_0(z)+\sum_{i=1}^m\psi_i(z)|z-z_i|^2,\tag 4.5
$$
 where
$\psi_0(z)=1$ everywhere except the disks $B(z_i,r)$ centered at
 $z_i$ of a small radius $r$, and $\psi_0(z)=0$ in $B(z_i,r/2)$;
for $i>0$,
$\psi_i$ is a nonnegative function equal to
$1$ in $B(z_i,r/2)$ and to $0$ outside $B(z_i,r)$
(of course, all $\psi_i$ have to be smooth).

{}From the definition of $\psi$ it follows that the function
$u=\d^2\psi/\psi$ defined a priori in $M\setminus\{z_1,...,z_n\}$,
can be continued to the points $z_1,...,z_n$ (since it is
simply equal to zero in their neighborhoods). This implies that
$\psi$ is a solution of the equation $L\psi=0$, where $L=\d^2-u$.
Pick a vector $\bold f=(f_1,f_2)$ of solutions of this differential
equation with a nondegenerate Wronski matrix.
Then there exist unique holomorphic functions $c_1(z)$, $c_2(z)$
on the cylinder $\Sigma$ such that $\psi=c_1f_1+c_2f_2$,
and the vector-function $\bold c=(c_1,c_2)$
is a global holomorphic section of the holomorphic vector bundle
$\E_L$.

Let us show that this section vanishes
at the points $z_1,...,z_m$ and only at them, and these zeroes are simple.
Indeed, the vector $\bold F=\left(\matrix \psi\\ \d\psi\endmatrix\right)$
equals $\bold cW(\bold f)$, thus $\bold c=0$ iff $\bold F$ vanishes, and
the vanishing points of $\bold F$ are exactly $z_1,...,z_m$.
Also, in the neighborhood of $z_i$ one has $\bold F=
(z-z_i)\left(\matrix \bar z-\bar z_i\\ 1\endmatrix\right)$, which
shows that $z_i$ is a simple zero of $\bold c$.

It follows from the theory of holomorphic bundles that the presence of
a section $\bold c$ with the above properties guarantees
that $\E_L$ has a line subbundle $X$ of degree $m$
defined by the monodromy function $e^{2\pi im(z-z_0)}$.
The bundle $\Lambda^2\E_L$ is trivial
since the operator $L$ does not contain a first order term,
and hence the Wronskian (which is a section of $\Lambda^2\Cal E_L$)
is constant. This fact together
with Atiyah's classification theorem implies that
$\E_L$ is isomorphic to $X\oplus X^*$, which is the same
as $E(a,a^{-1},m)$, where $a=e^{-2\pi imz_0}=\prod_je^{2\pi i z_j}$. Since
the points $z_i$ could be chosen arbitrarily, one can get any value of
$a$.
\vskip .1in

Let us now describe the codimensions and adjacency of symplectic leaves for
$n=2$ (Case 2). It follows Theorem 5AB, Proposition 7AB, and Corollary 8AB
that it is enough to do it for vector bundles.

\proclaim{Proposition 10B}
(i) If $\E_L=E(a,b,m)$ then $\text{codim}(\Cal O_L)$ equals $2m+2$
if $m>0$, $2$ if $m=0$ and $a/b$ is not an integral power of $q$,
and $4$ if $m=0$ and $a/b=q^k$, $k\in\Z$.

(ii) If $E_L=F(a)$ then $\text{codim}(\Cal O_L)$ equals $2$.

(iii) For $n=2$, a symplectic leaf
$\Cal O_{L_1}$ is adjacent to $\Cal O_{L_2}$
iff $\Lambda^2\E_{L_1}=\Lambda^2\E_{L_2}$ and
$\text{codim}\Cal O_{L_1}<\text{codim}\Cal O_{L_2}$.
\endproclaim

\demo{Proof} (i) $E(a,b,m)=X_{a,m}\oplus X_{b,-m}$, where $X_{a,m}$,
is the line bundle described by the monodromy
function $ae^{2\pi imz}$.
Therefore, $E(a,b,m)\otimes E(a,b,m)^*=X_{1,0}\oplus X_{1,0}\oplus X_{a/b,2m}
\oplus X_{b/a,-2m}$. The number of linearly independent
holomorphic sections of this bundle
is 2 if $m=0$ and $a/b\ne q^k$, 4 if $m=0$ and $a/b=q^k$, and $2m+2$ if
$m\ne 0$, which proves (i).

It is also easy to see that $F(a)\otimes F(a)^*=X_{1,0}\oplus F_3(1)$,
where $F_3(1)$ is the vector bundle of rank $3$ whose monodromy matrix
is the $3\times 3$ Jordan cell with eigenvalue 1.
Therefore, the number of linearly independent
holomorphic sections of $F(a)\otimes F(a)^*$ is 2.
This settles (ii).

The ``only if'' part of statement (iii) is obvious.
The ``if'' part is proved by means of case by case analysis, as follows.

Clearly, it is enough to consider the case when $\Lambda^2\E_{L_i}$ is
a trivial line bundle.

The adjacency of $E(a_1,a_1^{-1},m)$ to $E(a_2,a_2^{-1},m-1)$ for any $a_i$
and $m>1$ is established by introducing the family of functions
$\psi^t=\psi+t^2\psi_1$, where $\psi$ and $\psi_1$ are defined in the
Section 4 (see formula (4.5)), and
considering the curve $L_t$ of operators $\d^2-u_t$ such that
$L_t\psi^t=0$ (i.e. $u_t=\d^2\psi^t/\psi^t$).
It is easy to see that the monodromy of $L_t$ for $t=0$ is of
the type $E(a_1,a_1^{-1},m)$, and for $t\ne 0$ of the type
$E(a_2,a_2^{-1},m-1)$.

The same construction for $m=1$ demonstrates the adjacency of
$E(a,a^{-1},1)$ to $F(1)$.

To demonstrate that $E(a_1,a_1^{-1},1)$ is adjacent to
$E(a_2,a_2^{-1},0)$ when $a_2^2\ne q^k$, $k\in \Z$,
one should consider the above construction with a
minor modification: the function $\psi$ should be
a function on the cylinder $\Sigma$
given by formula (4.5) in which $\psi_i$ satisfy the condition
$\psi_i(z+\tau)=a\psi_i(z)$, $0\le i\le m$ and are chosen in such a
way that $\psi$ has properties (i) and (ii).

Finally, the adjacency of $E(1,1,0)$ to $F(1)$ is established as
follows. Let $\psi$ be any nonvanishing function on the elliptic
curve. Let $u=\d^2\psi/\psi$. Consider
the differential operator $L=\d^2-u$. It is easy to show that
the monodromy matrix of this operator defines the bundle $E(1,1,0)$
if $\int_{M}\biggl(\frac{1}{\psi}\biggr)^2\omega=0$, and the bundle $F(1)$ if
this
integral is nonzero. Therefore, if $\psi^t$ is a family of
nonvanishing functions such that
$\int_{M}\biggl(\frac{1}{\psi^t}\biggr)^2\omega=0$
if and only if $t=0$ then the corresponding operators
$L_t=\d^2-\d^2\psi^t/\psi^t$ have monodromy $F(1)$
if $t\ne 0$ and $E(1,1,0)$ if $t=0$.

The rest of adjacencies follow from the fact that adjacency
is a partial order.
$\square$\enddemo

{\bf Remark. } Observe an interesting
feature of the affine GD bracket which was
not present in the finite dimensional case: in Case 2
the codimensions of symplectic leaves, though all finite,
can be arbitrarily large, whereas in Case 1 they are bounded from
above by $n^2=\text{dim}GL_n$. However, the conjugacy classes labeling
all symplectic leaves of
codimension $>n^2$ stay away from the $(\text{Id},\tau)\in \cG$, by
virtue of Observation in Section 3.

\vskip .1in
{\bf 5. Classification of symplectic leaves with a given monodromy. }
\vskip .1in

In this section we will address the question of classification of
symplectic leaves with a given monodromy, or, equivalently,
the problem of finding discrete invariants of symplectic leaves.

In Case 1 this problem was studied in \cite{OK},
and it was shown that it is equivalent to the problem of
homotopy classification of quasiperiodic nondegenerate curves.

\proclaim{Definition 6A}\cite{OK} A quasiperiodic nondegenerate curve
(QN curve) in $k^n$ with monodromy matrix $R\in GL_n(k)$
 is a smooth function $\gamma(x)=(\gamma_1(x),...,\gamma_n(x))$
on the real line with two properties:

(i) quasiperiodicity: $\gamma(x+1)=\gamma(x)R$;

(ii) no degeneration points of the Wronski matrix: the vectors
$\gamma,\gamma',...,\gamma^{(n-1)}$ are linearly independent for any
$x$, and form a right-handed basis of $\R^n$ if $k=\R$.

Two QN curves are called homotopic if one of them can be deformed
into the other in such a way that all the intermediate curves are QN curves
with monodromy $R$.
\endproclaim

{\bf Remark.} In the case $k=\Bbb R$, we consider only curves with
a positive Wronskian (right-handed curves).

It is easy to show that for any $R$ there exists a QN curve
$\gamma$ with monodromy matrix $R$ such that $\gamma^{(j)}(0)=e_j$,
where $e_j$ is the vector $(0,...,0,1,0,...,0)$ with $1$ at the $j$-th place
(In the real case, it follows from \cite{S}; in the complex case,
take any quasiperiodic curve with monodromy $R$, then perturb it if
necessary to ensure the nondegeneracy of the Wronski matrix -- cf. the
proof of Proposition 9A). It is also routine to
prove that any QN curve with monodromy $R$ can be deformed into
one with $\gamma^{(j)}(0)=e_j$ inside the class of QN curves with
monodromy $R$. Therefore, considering homotopies of QN curves, we may
assume that $\gamma^{(j)}(0)=e_j$.

Besides smooth ($C^{\infty}$) QN curves, it is useful to consider
$C^{n-1}$-QN curves. Such curves are very easy to construct:
take any smooth curve $\gamma:[0,1]\to k^n$ which has no degeneration points
(see Definition 6A), define the monodromy matrix by
$\gamma^{(i)}(1)=\gamma^{(i)}(0)R=e_iR$, $1\le i\le n$, and
extend the function $\gamma$ to the entire real line
by setting $\gamma(x+n)=\gamma(x)R^n$. In terms of homotopy
properties, there is no difference between $C^{n-1}$ and $C^{\infty}$
QN curves, since every $C^{n-1}$ QN curve can be approximated by a
smooth one as closely as desired. Therefore, from now on we
deal with $C^{n-1}$ QN curves unless otherwise specified.

We will also consider projections of QN curves in $k^n$ to the
projective space $kP^{n-1}$.

\proclaim{Definition 7A} A curve $\hat\gamma: \R\to kP^{n-1}$
is called a quasiperiodic nonflattening curve (a QNF curve)
with monodromy $R\in GL_n(k)$ (for $k=\R$ we require $\text{det}R>0$)
if it satisfies the conditions

(i) $\hat\gamma(x+1)=\hat\gamma(x)R$
($\hat\gamma R$ denotes the result of the
action of the linear transformation $R$ on
the point $\hat\gamma$ on the projective space);

(ii)
the vectors $\hat\gamma'(x),...,\hat\gamma^{(n-1)}(x)$
are linearly independent at every point $x$;

and

(iii) in the case $k=\R$, there exists a lifting
$\tilde\gamma$ of $\hat\gamma$ to
the sphere $S^{n-1}$ such that
the vectors $\tilde\gamma',...,\tilde\gamma^{(n-1)}$
form a right-handed basis of the tangent space to the sphere at every
point of the curve $\tilde\gamma$.

Two QNF curves with the same monodromy $R$
are called homotopic if one of them can be deformed
into the other without leaving the class of
QNF curves.
\endproclaim

As in the case of QN curves, for QNF curves
we may assume that $\hat\gamma(0)$ is the line
generated by $e_1$, and the vectors $\hat\gamma',...,\hat\gamma^{(n-1)}$
are equal to the projections of $e_2,...,e_n$. As before, this does
not cause any loss of generality.

\proclaim{Proposition 11A}\cite{OK} Let $\hat\gamma$ be the image of a
QN curve $\gamma$ under the canonical projection $k^n\to kP^{n-1}$.
Then

(i) $\hat\gamma$ is a QNF curve;

(ii) any QNF
curve $\hat\gamma$ on $kP^{n-1}$
is a projection of a QN curve $\gamma$.
\endproclaim

\demo{Proof} The proof is straightforward.
$\square$\enddemo

\proclaim{Proposition 12A}\cite{OK}
Symplectic leaves of the $GL_n$-GD bracket whose monodromy is the
conjugacy class of $R$ in $GL_n(k)$ are in one-to one correspondence with
homotopy classes of quasiperiodic nondegenerate curves with monodromy $R$.
\endproclaim

\demo{Proof} Let $L$ be a differential operator. We can assign
a QN curve to $L$ by considering the fundamental
system of solutions $\bold f(x)=(f_1(x),...,f_n(x))$ to the equation
$L\phi=0$. It is easy to see that two such curves are homotopic as QN
curves (or, equivalently, their projections to $kP^{n-1}$ are
homotopic as QNF curves) iff the differential operators
from which they originated are from the same symplectic leaf.
$\square$\enddemo

Proposition 12A reduces the problem of finding discrete invariants
of symplectic leaves to a topological problem.

For $k=\R$ and general $n$ this
topological problem turns out to be difficult. It is solved
only for $n=2$ (where this problem is equivalent to classification of
projective structures on the circle \cite{Ku}, of Hill's operators
\cite{LP}, or coadjoint orbits of the Virasoro algebra
\cite{Ki},\cite{Se}),
for $n=3$ \cite {KS},and for any $n$ in case $R=\text{Id}$ \cite{S}.

In the complex case ($k=\C$),
one can introduce an obvious topological invariant
of a QN curve  -- the winding number.

Let $\gamma$ be a QN curve in $\Bbb C^n$, and let
$w(x)$ be the Wronskian of $\gamma$: $w(x)= \text{det}(\gamma_i^{(j)}(x))$.
Then $w(x)$ is quasiperiodic: $w(x+1)=rw(x)$, where $r=\text{det}(R)$.
Therefore, one can define the winding number
$$
\nu(\gamma)=\frac{1}{2\pi
i}\biggl(\int_0^1\frac{dw(x)}{w(x)}-\log r\biggr),\tag 5.1
$$
where $\log r$ is a fixed value of the logarithm of $r$. It is obvious
that $\nu(\gamma)$ is an integer and that it is invariant under
homotopy of QN curves.

It should be mentioned that the winding invariant is a feature of the
$GL_n(\C)$-Gelfand-Dickey consideration. For the $SL_n(\C)$-counterpart,
this invariant is trivial.

Let us show that the winding number can take any integer value.
Clearly, it is enough to show that there exists
a closed QN curve $\gamma$ (i.e. a QN curve
with monodromy $R=\text{Id}$) having winding number 1:
then for any other QN curve one can combine it with sufficiently many
copies of $\gamma$ (or reversed $\gamma$) at one period to ensure
that the winding number is as desired. The curve $\gamma$ can be,
for example, given by the formula $\gamma(x)=(e^{2\pi
im_1x},...,e^{2\pi im_nx})$. This curve is QN iff $m_j$ are all
distinct, and its winding number is $m_1+\dots+m_n$, so it can
be made equal to 1 if desired.

Let us now discuss Case 2 (elliptic curve). In this case the geometric notion
corresponding to the problem of finding discrete invariants of
symplectic leaves
is the notion of a {\it quasiperiodic $\d$-nondegenerate tube}.

\proclaim{Definition 6B} A quasiperiodic $\d$-nondegenerate tube
(a QN tube) in $\C^n$ with monodromy matrix $R(z)\in LGL_n(\C)_0$
 is a smooth function $\gamma(z)=(\gamma_1(z),...,\gamma_n(z))$
on the cylinder $\Sigma$
with two properties:

(i) quasiperiodicity: $\gamma(z+\tau)=\gamma(z)R(z)$;

(ii) no $\d$-degeneration points of the Wronski matrix: the vectors
$\gamma,\d\gamma,...,\d^{n-1}\gamma$ are linearly independent over
$\C$ for any
$z$.

Two QN tubes are called homotopic if one of them can be deformed
into the other so that all the intermediate tubes are QN tubes
with monodromy $R(z)$.
\endproclaim

As before, we will also consider projections of QN tubes in $\C^n$ to the
projective space $\C P^{n-1}$.

\proclaim{Definition 7B} A tube $\hat\gamma: \Sigma\to \C P^{n-1}$
is called a quasiperiodic $\d$-nonflattening tube (a QNF tube)
with monodromy $R(z)\in L\G_0$ if $\hat\gamma(z+1)=\hat\gamma(z)R(z)$, and
the vectors $\d\hat\gamma,...,\d^{n-1}\hat\gamma$
are linearly independent at every point $z$.
Two QNF tubes with the same monodromy $R(z)$
are called homotopic if one of them can be deformed
into the other without leaving the class of
QNF tubes.
\endproclaim

\proclaim{Proposition 11B} Let $\hat\gamma$ be the image of a
QN tube $\gamma$ under the canonical projection $\C^n\to \C P^{n-1}$.
Then

(i) $\hat\gamma$ is a QNF tube;

(ii) any QNF
tube $\hat\gamma$ on $\C P^{n-1}$
is a projection of a QN tube $\gamma$.
\endproclaim

\demo{Proof} The proof is straightforward, like in Case 1
$\square$\enddemo

\proclaim{Proposition 12B}
Symplectic leaves of the affine $GL_n$-GD bracket whose monodromy is the
conjugacy class of $R(z)$ are in one-to one correspondence with
homotopy classes of quasiperiodic nondegenerate tubes with monodromy $R(z)$.
\endproclaim

\demo{Proof} Let $L$ be a differential operator. We can assign
a QN tube to $L$ by considering a
system of solutions $\bold f(z)=(f_1(z),...,f_n(z))$ to the equation
$L\phi=0$. It is easy to see that two such tubes are homotopic as QN
tubes iff the differential operators
from which they originated are from the same symplectic leaf.
$\square$\enddemo

Similarly to Case 1,
Proposition 12B reduces the problem of finding discrete invariants
of symplectic leaves to a topological problem.
Unfortunately, we do not have a complete solution to this problem even
for $n=2$. However, as in Case 1,
we can construct some topological invariants.

One can introduce two obvious topological invariants
of a QN tube in $\C^n$ -- the winding invariants.
Let $\gamma$ be a QN tube, and let
$w(z)$ be the Wronskian of $\gamma$: $w(z)= \text{det}(\d^j\gamma_i(z))$.
Then one can define the winding number
$$
\nu_1(\gamma)=\frac{1}{2\pi i}\int_{0}^1\frac{dw(z)}{w(z)}.\tag 5.2
$$
Also, since $w(z)$ is quasiperiodic:
$w(z+\tau)=r(z)w(z)$, where $r(z)=\text{det}(R(z))$, one can
 define the second winding number as follows:
$$
\nu_2(\gamma)=\frac{1}{2\pi
i}\biggl(\int_0^1 d_y w(x+y\tau)-
{\log r(x)}\biggr),\tag 5.3
$$
where $\log r$ is a fixed branch of the logarithm of $r$
(clearly, (5.3) is independent of $x$). It is obvious
that $\nu_i(\gamma)$, $i=1,2$, are integers and that they are
invariant under
homotopy of QN tubes.

If $R(z)=\text{Id}$,
we can show that the winding numbers can take any integer values.
The tube $\gamma$ can be,
for example, given by the formula $\gamma(z)=(e^{2\pi
i(m_1x+p_1y)},...,e^{2\pi i(m_nx+p_ny)})$. This tube is QN iff
$(m_j,p_j)$
are all
distinct, and its winding numbers are $\nu_1=m_1+\dots+m_n$, $\nu_2=
p_1+\dots+p_n$, so they can
be made equal to any desired numbers.

Let us consider the case $R(z)=\text{Id}$ in more detail.
In this case the QN (QNF) tubes we are dealing with are
closed, i.e. they are smooth maps from $M$ to $\C^n$ ($\C
P^{n-1}$), and the geometric reformulation of the problem
of finding discrete invariants of symplectic leaves is especially
elegant.

\proclaim{Proposition 13B} Symplectic leaves of the affine $GL_n$-GD
bracket whose monodromy is the identity (=the trivial bundle)
are in one-to-one correspondence with homotopy classes of maps
from an elliptic curve to $\C^n$ whose Wronski matrix is
everywhere nondegenerate.
\endproclaim

\demo{Proof} This statement follows from Proposition 12B.
$\square$\enddemo

It is clear that if we multiply a QN tube by a nonvanishing function,
we will get another QN tube; these two QN tubes will project
to the same QNF tube on the projective space. It is also clear
what happens to the winding numbers when a QN tube $\gamma$ is multiplied by a
nonvanishing function $\phi$:
$\nu_i(\phi\gamma)=\nu_i(\gamma)+n\nu_i(\phi)$.
Therefore, if we identify QN tubes which differ
by a nonvanishing function (i.e.  if we consider QNF tubes),
the winding numbers are defined only modulo $n$, i.e.  they are
elements of $\Z/n\Z$.

This reasoning shows that Proposition 13B can be formulated as
follows:

\proclaim{Proposition 13B*}
Symplectic leaves of the affine $GL_n$-GD
bracket whose monodromy is the trivial bundle and whose winding
numbers are $\nu_1$, $\nu_2$
are in one-to-one correspondence with homotopy classes of maps $f$
from an elliptic curve to $\C P^n$ such that
the vectors $\d f,...,\d^{n-1} f$ are everywhere linearly independent,
and $\nu_i(f)=\nu_i\text{ mod }n$.
\endproclaim

{\bf Example: n=2. } In this case the problem
of classification of symplectic leaves reduces to the problem of
homotopy classification of nowhere holomorphic maps.

\proclaim{Definition 8B} Let $M$ be an elliptic curve.
A map $f:M\to\C P^1$ is called nowhere holomorphic if
$\d f$ is not equal to zero at any point of $M$.
\endproclaim

It is easy to show that every nowhere holomorphic map has degree zero
since it always comes from a map $\gamma: M\to \C^2$.

The winding numbers of a nowhere holomorphic map are elements of
$\Z/2\Z$. They can be defined as follows.
Consider the map $z\to \d f(z)/|\d f(z)|$ from the torus $M$ to
the space of unit tangent vectors to $\C P^1$.
Since this space is diffeomorphic to $\R P^3$,
the induced map of fundamental groups maps $\Z\oplus \Z$ to $\Z/2\Z$.
Then the winding numbers $\nu_1,\nu_2$ are just the images of the
generators $(1,0)$ and $(0,1)$ of $\Z\oplus\Z$ in $\Z/2\Z$.

Thus, the question whether a  symplectic leaf
is uniquely
defined by its winding numbers is equivalent to the following question:

\proclaim{Open Question} Is it true that two nowhere holomorphic
maps are homotopic in the class of nowhere holomorphic maps
if and only if their winding numbers are the same?
\endproclaim

{\bf Remark. }
It is easy to show that if the images of both maps are not the entire
$\C P^1$, the answer to this question is positive.
Therefore, to settle this question, it would be enough to show
that any nowhere holomorphic map can be deformed into another one
which misses at least one point in $\C P^1$.

\end
\Refs

\ref\by [A] Adler, M.\paper On a trace functional for formal
pseudo-differential operators and the symplectic structure of the
Korteweg-Devries equations\jour Inv. Math.\vol 50\pages 219-248\yr
1979\endref

\ref\by [AB] Atiyah, M., and Bott, R.\paper The Yang-Mills equations
over Riemann surfaces, Philos. Trans.Roy. Soc. London \vol A 308
\pages 523-615\yr 1982\endref

\ref\by [At] Atiyah, M.\paper Vector bundles over an elliptic curve
\jour Proc. Lond. Math. Soc. \vol 7\pages 414-452\yr 1957\endref

\ref\by [DS] Drinfeld, V.G., and Sokolov, V.V.\paper Lie algebras and
equations of the Korteweg-de-Vries type\jour J. Soviet Math.\vol
30\pages 1975-2036\yr 1985\endref

\ref\by [EF] Etingof, P.I., and Frenkel, I.B.\paper Central extensions
of current groups in two dimensions\jour hep-th 9303047\yr 1993\endref

\ref\by [F] Frenkel, I.B.\paper Orbital theory for affine Lie algebras
\jour Inventiones Mathematicae\vol 77\pages 301-352\yr 1984\endref

\ref\by [GD] Gelfand, I.M., and Dickey, L.A.\paper A family of
Hamiltonian structures related to integrable nonlinear differentail
equations, Preprint, 1978, English translation in: I.M.Gelfand,
Collected papers, Vol.1, Gindikin S.G.,et. al.
(eds.) Berlin, Heidelberg, New York: Springer 1987\endref

\ref \key Ki
\by A. A. Kirillov
\paper Infinite-dimensional Lie groups:  their orbits, invariants and
representations.  Geometry of moments
\jour in Lect. Notes in Math., Vol\.~970 (1982), Springer--Verlag
\pages 101--123
\endref

\ref\by [KS] Khesin, B.A., and Shapiro, B.Z.\paper Nondegenerate
curves on $S^2$ and orbit classification of the Zamolodchikov algebra
\jour Comm. Math. Phys.\vol 145\pages 357-362\yr 1992\endref

\ref \key Ku
\by N. H. Kuiper
\paper Locally projective spaces of dimension one
\jour Michigan Math. J., Vol\.~2 (1953--1954), No\.~2 \pages 95--97
\endref

\ref \key LP
\by V. P. Lazutkin and T. F. Pankratova
\paper Normal forms and versal deformations for the Hill's equations
\jour Funct. Anal. and Appl. {\bf 9} (1975), No\.~4 \pages 41--48
\endref

\ref \key O
\by V. Yu. Ovsienko
\paper Classification of linear differential equations of third order
and symplectic leaves of the Gel'fand--Dikii bracket
\jour Math. Notes {\bf 47} (1990), No\.~5 \pages 62--69
\endref

\ref\by [OK] Ovsienko, V.Yu., and Khesin, B.A.\paper Simplectic leaves
of the Gelfand-Dickey brackets and homotopy classes of nondegenerate
curves
\jour Funct. Anal. Appl.\vol 24(1)\pages 33-40\yr 1990\endref

\ref\by [PS] Pressley, A., and Segal, G.\book Loop groups\publ
Clarendon
Press\publaddr
Oxford \yr 1986\endref

\ref\by [RS] Reiman, A.G., and Semenov-Tian-Shansky, M.A.\paper
Lie algebras and nonlinear partial differential equations \jour Soviet
Math. Doklady\vol 21\pages 630\yr 1980\endref

\ref \key S
\by M. Z. Shapiro
\paper Topology of the space of nondegenerate curves
\jour Funct. Anal.Appl.\vol 26(3)\pages 227-229\
\yr 1991
\endref

\ref \key Se
\by G. Segal
\paper Unitary representations of some infinite dimensional groups
\jour Comm. Math. Phys. {\bf 80} (1981), No\.~3 \pages 301--342
\endref
